\documentclass[sigconf, usenames, dvipsnames]{acmart}

\AtBeginDocument{%
  \providecommand\BibTeX{{%
    \normalfont B\kern-0.5em{\scshape i\kern-0.25em b}\kern-0.8em\TeX}}}

\usepackage{comment}
\usepackage{subfig}
\usepackage{xspace}
\usepackage{color,soul}
\setstcolor{lightgray} 
\usepackage{textgreek}

\copyrightyear{2024}
\acmYear{2024}
\setcopyright{acmlicensed}\acmConference[arXiv]{}{May, 2024}{}

\begin{document}


\newcommand{\system}{RealityChat\xspace}

\newcommand{\todo}[1]{{\color{purple}{[#1]}}}
\newcommand{\ryo}[1]{\textcolor{purple}{[Ryo:#1]}}
\newcommand{\ww}[1]{\textcolor{orange}{[Wes:#1]}}

\newcommand{\completed}[1]{\textcolor{lightgray}{\st{#1}}}

\title{Augmented Conversation with Embedded Speech-Driven On-the-Fly Referencing in AR}

\author{Shivesh Jadon}
\affiliation{%
  \institution{University of Calgary}
  \city{Calgary}
  \country{Canada}}
\email{shivesh.jadon@ucalgary.ca}

\author{Mehrad Faridan}
\affiliation{%
  \institution{University of Calgary}
  \city{Calgary}
  \country{Canada}}
\email{mehrad.faridan1@ucalgary.ca}

\author{Edward Mah}
\affiliation{%
  \institution{University of Calgary}
  \city{Calgary}
  \country{Canada}}
\email{edward.mah@ucalgary.ca}

\author{Rajan Vaish}
\affiliation{%
  \institution{Snap Research, Easel AI}
  \city{Los Angeles}
  \country{United States}}
\email{vaish.rajan@gmail.com}

\author{Wesley Willett}
\affiliation{%
  \institution{University of Calgary}
  \city{Calgary}
  \country{Canada}}
\email{wesley.willett@ucalgary.ca}

\author{Ryo Suzuki}
\affiliation{%
  \institution{University of Calgary}
  \city{Calgary}
  \country{Canada}}
\email{ryo.suzuki@ucalgary.ca}

\renewcommand{\shortauthors}{Jadon, et al.}

\newcommand{\subsub}[1]{{\vspace{0.3cm}\noindent\textbf{\textit{#1:}}}}

\begin{abstract}
This paper introduces the concept of \textbf{\textit{augmented conversation}}, which aims to support co-located in-person conversations via embedded \textit{speech-driven on-the-fly referencing} in augmented reality (AR).
Today computing technologies like smartphones allow quick access to a variety of references during the conversation. However, these tools often create distractions, reducing eye contact and forcing users to focus their attention on phone screens and manually enter keywords to access relevant information. 
In contrast, AR-based on-the-fly referencing provides relevant visual references in real-time, based on keywords extracted automatically from the spoken conversation. 
By embedding these visual references in AR around the conversation partner, augmented conversation reduces distraction and friction, allowing users to maintain eye contact and supporting more natural social interactions. 
To demonstrate this concept, we developed \system, a Hololens-based interface that leverages real-time speech recognition, natural language processing and gaze-based interactions for on-the-fly embedded visual referencing.
In this paper, we explore the design space of visual referencing for conversations, and describe our our implementation --- building on seven design guidelines identified through a user-centered design process. 
An initial user study confirms that our system decreases distraction and friction in conversations compared to smartphone searches, while providing highly useful and relevant information.
\end{abstract}

\ccsdesc[500]{Human-centered computing~Mixed / augmented reality}

\keywords{augmented reality; mixed reality; natural language processing; speech recognition; keyword extraction; embedded visual referencest}

\begin{teaserfigure}
\centering
\includegraphics[width=1\textwidth]{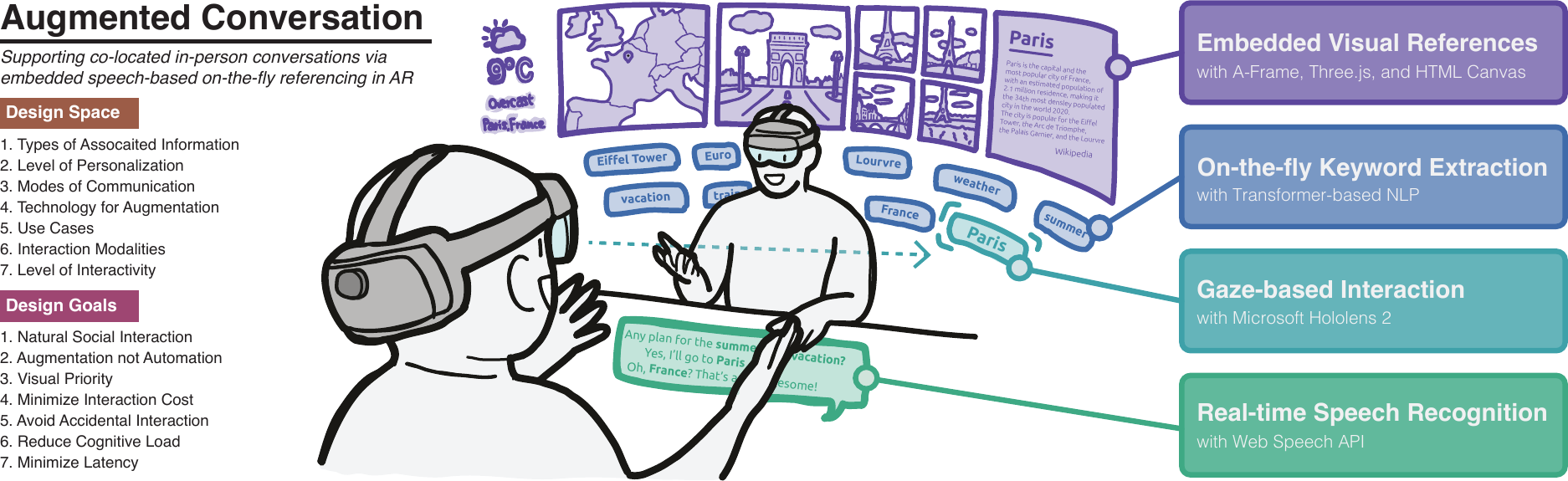}
\caption{On-the-fly conversational support through interactive augmented reality application. Our implementation transforms text from real-time speech recognition into visual overlays including interactive keywords and embedded visual references like maps, calendars, weather forecasts, photo galleries, Wikipedia articles, and search results. Individuals can use lightweight gaze and dwell interactions to select keywords and reveal additional details while remaining engaged in the conversation.}
\label{fig:teaser}
\end{teaserfigure}

\maketitle


\begin{figure*}[h!]
\centering
\resizebox{\textwidth}{!}{
\includegraphics[height=1\textwidth]{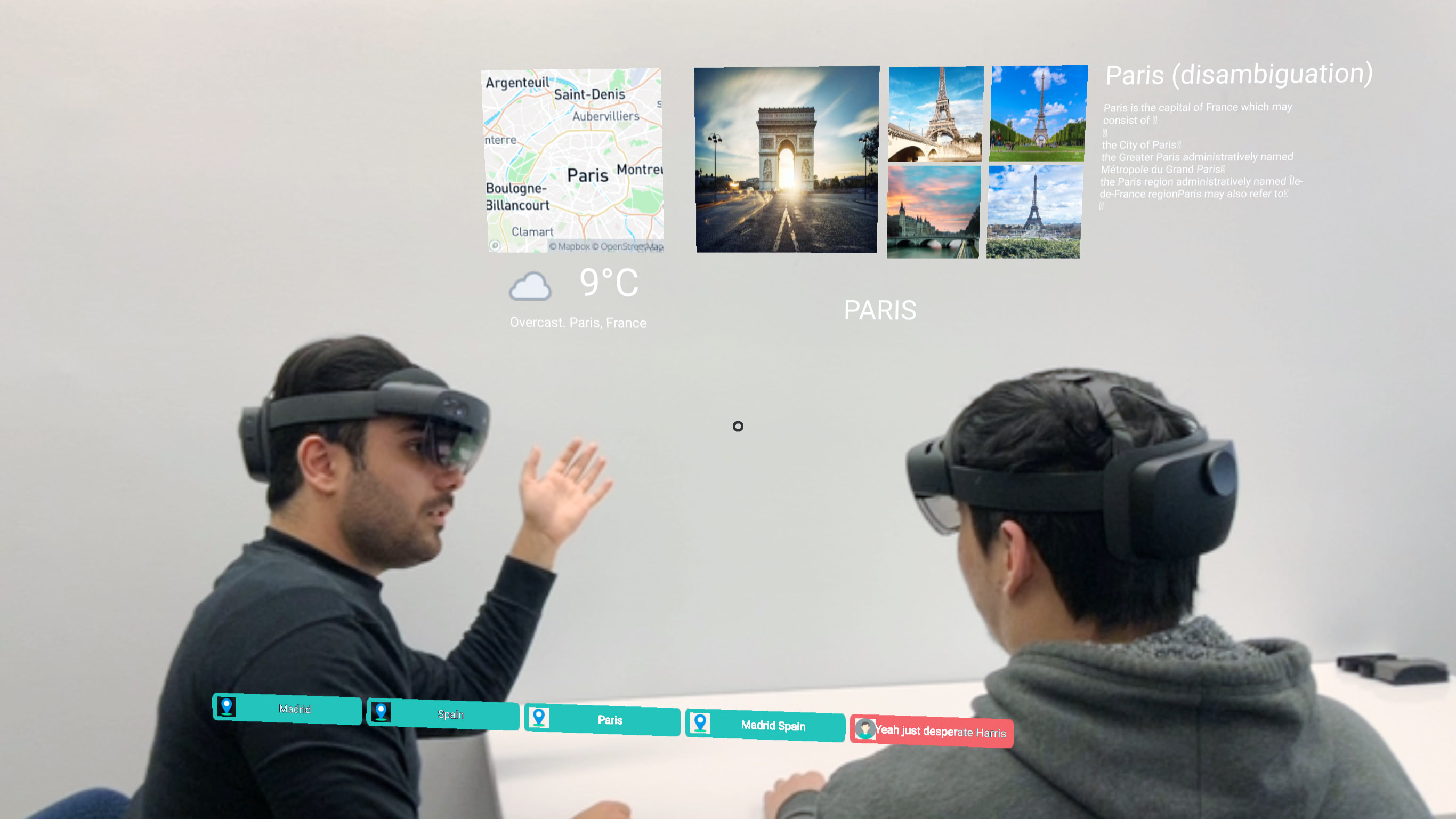}
\hspace{0.1cm}
\includegraphics[height=1\textwidth]{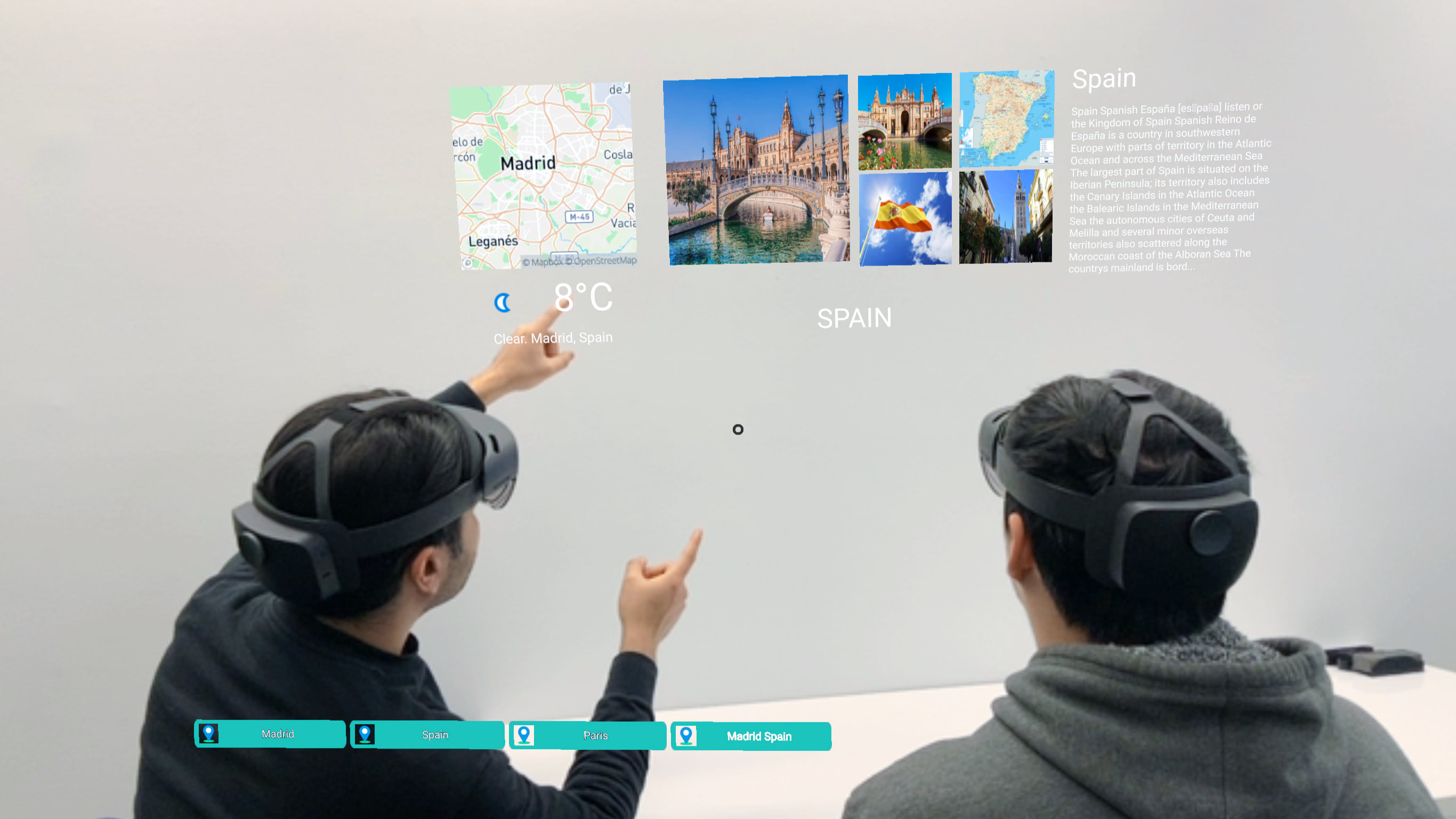}
\hspace{0.1cm}
\includegraphics[height=1\textwidth]{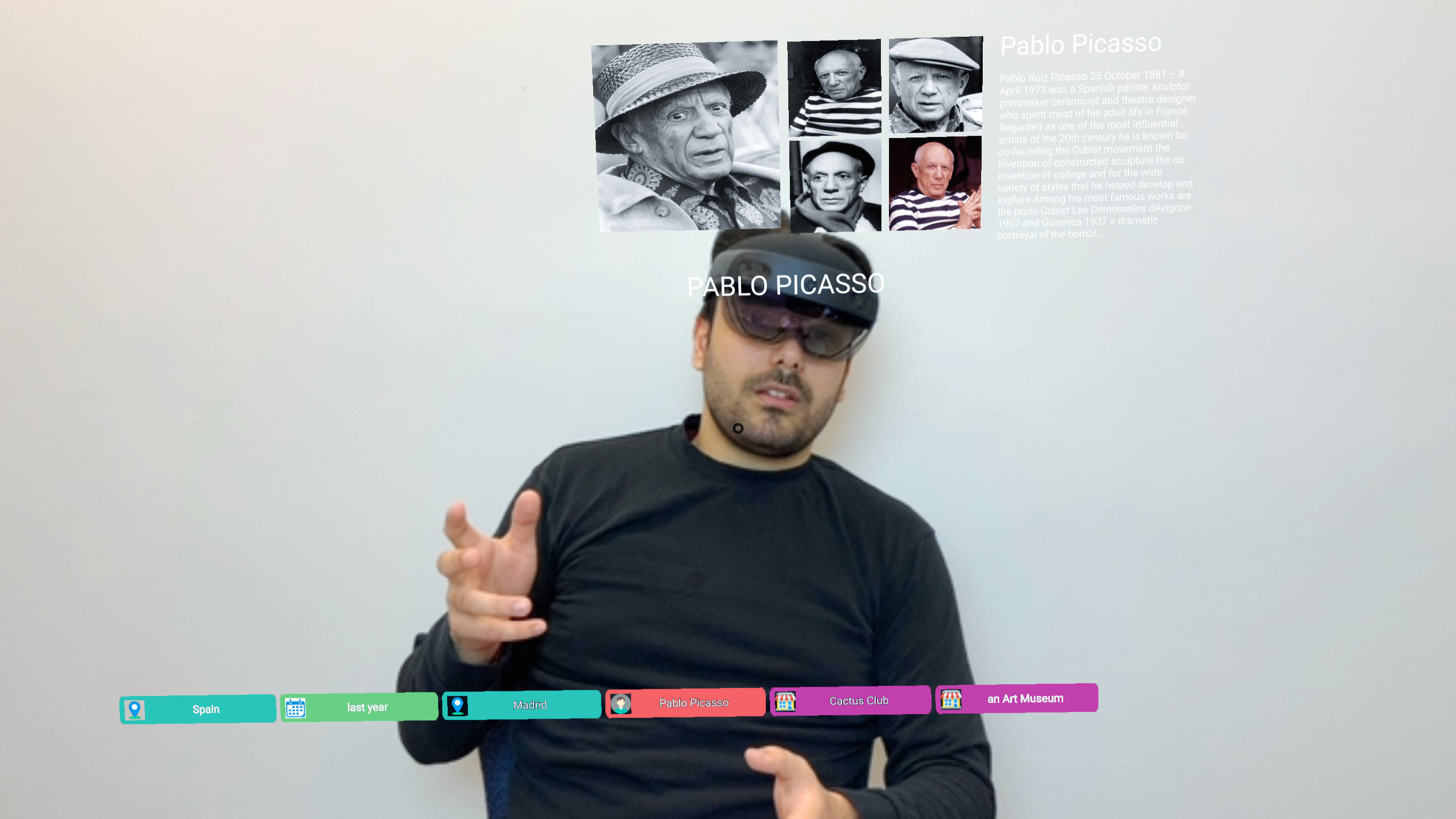}
}
\caption{On-the-fly conversational support through an interactive augmented reality (AR) application. (Left and Middle) Users interact with an AR interface that transforms real-time speech into visual overlays, displaying information about Paris, including maps, weather forecasts, and landmarks. (Right) A user explores details about Pablo Picasso through an overlay featuring images and textual information. This AR-based system exemplifies the concept of augmented conversation by providing real-time visual references based on spoken conversation keywords.}
\label{fig:}
\end{figure*}

\section{Introduction}
Today, computing technologies like smartphones have both positive and negative impacts on co-located social conversations.
On one hand, these devices allow quick access to a huge variety of information that can complement ongoing discussions.
For example, we can quickly search for restaurant locations and ratings while making lunch plans, rapidly share photos or videos of recent trips when talking to friends, or quickly examine news or Wikipedia articles when unfamiliar topics or questions arise in conversations.
Most smartphone users rely on these kinds of \textit{context-related on-the-fly referencing} every day~\cite{horrigan2015home}.
However, the use of smartphones also increases friction and distraction~\cite{dwyer2018smartphone}, leaving people unable to concentrate fully on conversations. 
Focusing on smartphone screens can reduce interpersonal eye contact, which may weaken the feeling of connection~\cite{yaziciouglu2017smartphone}.
Moreover, the presence of phones can distract people from their immediate social environment, potentially decreasing the enjoyment of social interactions~\cite{przybylski2013can, srivastava2005mobile}.

To address these challenges, we explore the design of technologies that can support everyday conversation without undermining our natural social interactions.
Specifically, we examine how augmented reality (AR) tools can \textit{reduce friction} by presenting information dynamically during conversations, eliminating the need for multi-step smartphone searches. Presenting information visually and spatially in AR could also \textit{reduce distraction}, making it possible to examine and share references without disrupting eye contact or interrupting social interactions. 

While prior work (like M\"uller et al.'s CloudBits~\cite{muller2017cloudbits}) has examined initial opportunities for presenting information in AR during conversations, we provide a deeper exploration of the design space of AR conversation support systems. We also demonstrate and evaluate an end-to-end system that leverages real-time speech recognition, language processing, and gaze to support fluid visual exploration of relevant references during conversations.

To provide a better understanding of the space of on-the-fly conversation augmentation, we first outline the concept of augmented conversation and present a design space that unpacks choices including: types of associated information, levels of personalization, modes of communication, technology, conversation partners, interaction modalities and levels of interactivity. We then introduce \system{}, an augmented reality interface that leverages real-time speech recognition, natural language processing, and gaze-based interactive visual reference search in AR (Figure~\ref{fig:teaser}).
\system{} has the following key features: 
\textbf{1) Real-time speech recognition and transcription} lets the system recognize and transcribe the speech in real-time, letting users see the conversation unfold visually.
\textbf{2) On-the-fly keyword extraction and referencing} translates key words in the conversation into \textit{interactive} elements that provide a fast and fluid alternative to mid-conversation smartphone searches.
\textbf{3) Visual embedding of references in AR} situates interactive keywords, maps, photos, and other visual references in the space immediately around the conversation partner, allowing users to quickly examine references without losing eye contact.
\textbf{4) Gaze-based interaction} allows users to interact with visual elements without the use of gesture or speech commands which might disrupt ongoing conversations.
Our system uses the Microsoft Hololens 2 with modern WebXR tools including A-Frame, Three.js, and React.js. We also use the Web Speech API for real-time speech recognition, a Transformer-based neural network NLP pipeline for real-time keyword extraction, and various web-based APIs (Google Maps, Google Images, Wikipedia, etc.) for context-driven reference searchs.

We evaluate the effectiveness of the interface with a usability study with thirteen participants.
Our overall findings indicate that, although participants were split on whether system was distracting or not, most of them agreed that it was intuitive ($\mu$=5, $\sigma$=1.15), reduced friction and provided better, on-the-fly referencing.  Majority of the participants agreed that system features were helpful in reducing the distraction, especially pictures, map, and Wikipedia articles. The participants provided a very positive feedback with all most all the participants agreeing that the referencing features were very helpful and directly relevant in the conversation. However, most of the participants preferred visual features over textual information. In terms of
the design, the participants generally think the interface is well-designed, but needs improvement in visibility, and reference scale and positioning. However, almost all participants found the system interface to be intuitive and as natural as daily conversations ($\mu$=5, $\sigma$=1.15).    
When asked about the usability of RealityChat in other social settings, participants were highly optimistic about its usefulness in educational (taking lecture notes, attending conferences, group studying, etc.) and professional settings (attending meetings, presenting, brainstorming sessions, etc.). They noted that keywords could help in creating an outline in lectures and conferences (P8), breakdown subject matter in steps for note taking (P9), help naturally search information in conversation (P2), easily take notes while listening to lectures (P4), and provide visual referencing to spoken words in lectures (P11). 

For future work, we also discuss the realm of improvement and trade-off of the current system (e.g., different use cases, privacy issues, personalized and context-aware information, etc) based on the participants' feedback.

The paper makes the following contributions: 
\begin{enumerate}
\item A design space characterizing \textbf{\textit{augmented conversation}} for on-the-fly context-driven referencing in augmented reality.
\item \system{}, a proof-of-concept augmented conversation prototype that demonstrates the potential of real-time speech recognition, keyword extraction, visual embedding of references, and gaze-based interaction.
\item Results and insights from an initial study that highlight the benefits and limitations of these approaches.
\end{enumerate}

\section{Background and Related Work}
Our research builds on previous work exploring the smartphones and user distraction, as well as existing systems for augmenting conversation, and presenting live information in AR. 

\subsection{Smartphone Use and Distraction}
A growing body of research has shown that the use of smartphones in a social setting can negatively influence face-to-face interactions. While smartphone use can make individuals feel more productive by providing on-the-fly information that may be relevant, it often triggers more negative reactions from others.
Past work has shown that smartphone use can cause distraction from the original topic of the conversation~\cite{przybylski2013can}, isolate users from their surroundings and inhibit engagement in public settings~\cite{su2015third}.
\footnote{This isolation effect has been referred to as \textit{phubbing}~\cite{haigh2015stop}, derived from a portmanteau of phone and snubbing.}
There is also qualitative and quantitative evidence~\cite{book:2414330} showing that using smartphones during conversations may degrade the shared social experience~\cite{shah2003motivational}, reduce conversation quality~\cite{przybylski2013can}, and limit social interactions~\cite{custers2005positive}. 

\subsection{Context-Aware Conversational Assistance}
To help address these issues, researchers in the HCI, CSCW, and visualization communities have explored approaches for \textit{ambient} and \textit{seamless} conversation assistance and awareness.
For example, researchers have explored approaches for aiding participants during conversations~\cite{eddie2015our}, supporting information retrieval in conversations through shared displays~\cite{lundgren2013bursting, moser2016technology}, providing adaptive notifications~\cite{lopez2015managing}, and displaying topics for common interest~\cite{nguyen2015known}.
ClearBoard~\cite{ishii1992clearboard} also demonstrates how augmented communication can facilitate more seamless and natural collaboration by retaining eye-contact in remote collaboration settings.

Meanwhile, visualization tools like Bergstrom and Karahalios's Conversation Clocks~\cite{bergstrom2007conversation} and Kim et al.'s Meeting Mediator~\cite{kim2008meeting} have sought to use live meeting audio to provide sociometric signals to help participants manage turn-taking and group interaction during meetings. 
Other visualization tools (including examples like Conversation Clusters~\cite{bergstrom2009conversation}, Shi et al.'s MeetingVis~\cite{shi2018meetingvis}, and Aseniero et al.'s MeetCues~\cite{aseniero2020meetcues}) provide visual summaries of meetings, but --- due to processing and rendering challenges --- most work has tended to focus on post-meeting reviews rather than live access.

In co-located settings, researchers have explored context-aware conversational assistants through \textit{ambient voice search}~\cite{radeck2014towards} or \textit{zero query search}.
For example, Ambient Search~\cite{radeck2014towards} is a screen-based conversational assistant that displays and retrieves relevant documents based on the topic of conversation.
Similarly, zero-query search or just-in-time information retrieval~\cite{rhodes2000just} is a concept that provides information in nonintrusive manner based on the user's local context. 
This is becoming more available in commercial voice assistants, such as Google Now or Microsoft Cortana. 

However, most existing work has still focused on surfacing information on desktop or mobile displays~\cite{radeck2014towards, shi2018meetingvis, bergstrom2009conversation, aseniero2020meetcues, estg2019smart}. 
The advent of mainstream augmented reality hardware presents exciting opportunities to embed conversation aids into real world environments~\cite{muller2017cloudbits, monteiro2023teachable, rivu2020stare}.
Initial explorations of this approach, including M\"uller et al.'s CloudBits~\cite{muller2017cloudbits} and Rivu et al.'s StARe~\cite{rivu2020stare} have used Wizard-of-Oz approaches to highlight the potential of simple conversation aids in AR.
Within this space, CloudBits focuses primarily on visualization and interaction techniques for revealing and arranging relevant snippets of shared information in one-on-one conversation settings. Meanwhile StARe examines the potential for gaze-based interactions with AR content positioned in space around a conversation partner. We expand upon this work, examining a broader design space of conversation augmentations and demonstrating a real-time end-to-end system that leverages speech recognition, natural language processing, and gaze-based interactions to support on-the-fly conversation augmentation in AR. 

\subsection{Real-time Captioning in AR}
While our work features real-time speech recognition as a key component, AR-based real-time captioning itself is not new~\cite{liao2022realitytalk}.
However, most of the existing applications have focused on transcription from an accessibility perspective, providing automatically generating a closed-captions for deaf and hard-of-hearing (DHH) individuals~\cite{mosbah2006speech, peng2018speechbubbles, olwal2020wearable, tu2020conversational, miller2017use, schipper2017caption, findlater2019deaf, jain2018exploring, jain2015head, jain2018towards}.  
For example, Jain et al.~\cite{jain2018towards} have developed accessible AR conversation aids to support captioning, environment navigation, and attentional balance between speakers. Similarly, Peng et al.'s SpeechBubbles~\cite{peng2018speechbubbles}, Olwal et al.'s Wearable Subtitles~\cite{olwal2020wearable}, and Suemitsu et al.'s Caption Support System~\cite{suemitsu2015caption} surface transcribed speech in viewers' field of view, helping DHH individuals visually interpret conversations. 
On top of captioning, researchers in this space have also investigated related factors such as visibility, usability, glanceability, and placement of AR closed captions~\cite{schipper2017caption, jain2018exploring}. 
However, these AR-based speech tools have tended to focus on \textit{recognition and captioning}, with little additional support for interaction or information retrieval based on them. Our work examines opportunities for making these embedded transcripts \textit{interactive}, enabling and supporting access to a richer range of information to complement and extend the text. With the recent advent of large language models (LLMs), we believe the integration of AR and AI~\cite{suzuki2023xr} will become important for interactive conversation support in AR.

\subsection{Gaze-Based Interaction in AR}
Finally, a variety of recent projects have used head-mounted AR displays for context-aware task support~\cite{lindlbauer2019context}, providing adaptive information panels~\cite{dudley2021crowdsourcing} and showing visual references related to ongoing tasks~\cite{kim2019evaluating}. Gaze-based interaction is becoming an increasingly popular input modality for hands-free, in-situ, and unobtrusive interaction~\cite{pfeuffer2020empirical} making it a great method of interaction for the these kinds of information displays. 
For example, Lu et al.'s Glanceable AR~\cite{lu2020glanceable} uses gaze to access information via unique interactions including head-glance, gaze-summon, and eye-glance. ARtention~\cite{pfeuffer2021artention} explores a design space of AR gaze interaction across three dimensions: real vs. virtual, layers of content, and consumption vs. selection.
Systems like Shop-i~\cite{kim2015shop}, Looking for Info~\cite{piening2021looking}, and StARe~\cite{rivu2020stare} also highlight how a gaze-based interaction can be useful for on-demand information retrieval. Taking inspiration from this work, our system also uses gaze-based interactions to support in-situ and hands-free information access during live conversations. 


\section{Augmented Conversation: Concept}

We introduce the concept of \textbf{\textit{augmented conversation}} to describe new interfaces that allow people to enhance their in-person conversations through \textit{on-the-fly context-driven referencing}.
The core idea of this approach is a focus on \textbf{\textit{speech and attention-driven}} information retrieval, eschewing explicit keyword searches in favor of ambient presentation of references based on the content of the conversation and the use lightweight attention-based interactions to navigate and explore them. Augmented conversation interfaces like these can display useful contextual information during conversations, while allowing people to keep their focus primarily on the discussion at hand and it's social participants.

\begin{figure*}[h!]
\centering
\includegraphics[width=1\textwidth]{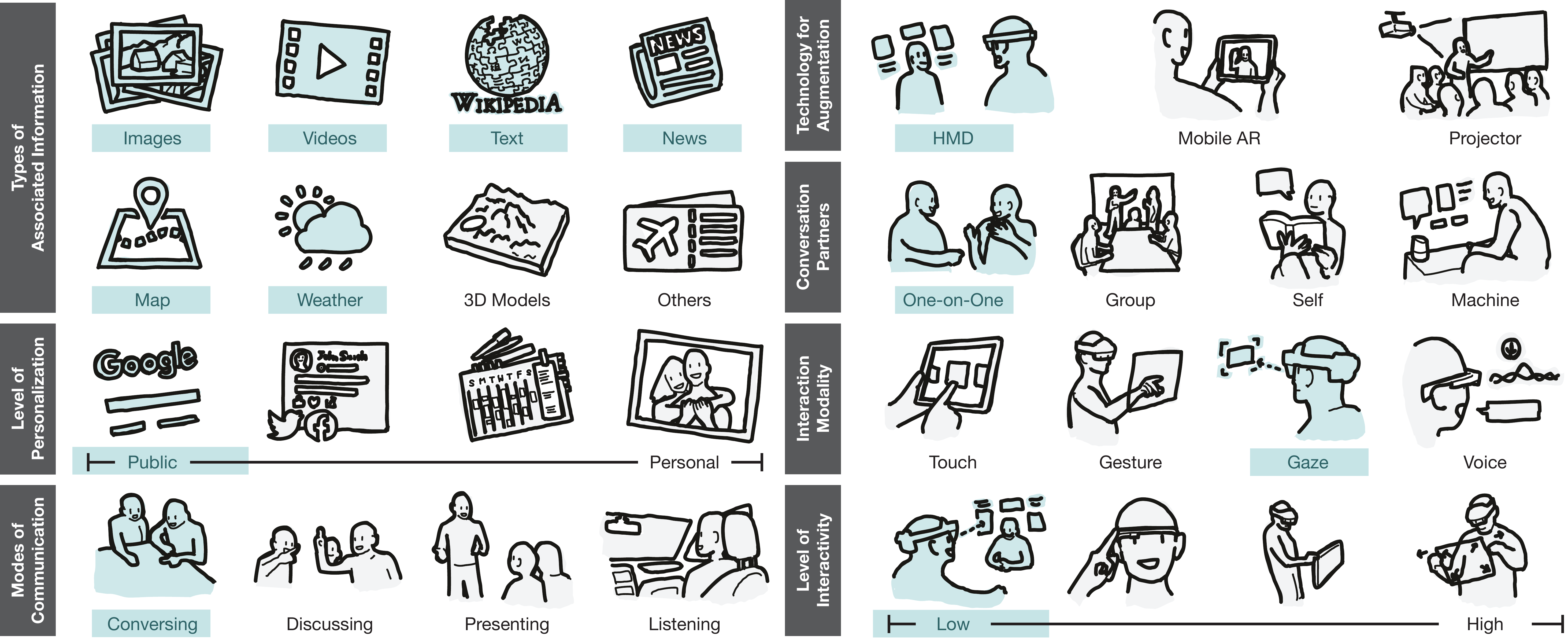}
\caption{The design space of augmented conversation approaches is large and diverse. Here we identify a variety of possible design options for augmented conversation spanning seven design dimensions. Design possibilities explored in our \system prototype are highlighted in \colorbox{teal!20}{\color{teal}{blue}}.}
\label{fig:designspace}
\end{figure*}

\subsection{Key Features}
As highlighted in these scenarios, our vision of augmented conversation encompasses several key features:

\subsub{1. Real-Time Speech Transcription}
Recognizing and transcribing speech in real-time to create a text record of the conversation.

\subsub{2. On-the-Fly Referencing}
Dynamically identifying key terms in the conversation and using them to automatically conduct on-the-fly searches for relevant content.

\subsub{3. Contextual Information Presentation}
Using contextual information about the scene and references to reduce clutter and highlight relevant information using appropriate representations such as images, maps, article snippets, etc. 

\subsub{4. Embedded Visual References}
Integrating visual references into the surrounding environment and/or around other speakers to help viewers maintain eye-contact and focus rather than staring at screens or mobile devices.

\subsub{5. Natural Interaction}
Relying on implicit and explicit natural interactions to maintain focus on the live conversation, rather than interacting with an interface. 


\subsection{Design Space}

As illustrated by the scenarios above, augmented conversation tools have the potential to take many different forms --- relying on a variety of hardware platforms, interactions, reference types, and visual representations to suit different kinds of conversations and environments. Here, we present an initial design space (Figure~\ref{fig:designspace}) that illustrates the diversity of possible augmented conversation approaches, as well as some promising opportunities.  

\subsub{D1. Types of Associated Information}
Augmented conversation has the potential to enhance conversations with many different types of associated information. These could include a variety of visual, spatial, and text-based content including images, videos, text description, news articles, charts, networks, 3D models, maps, and calendars, as well as other context-aware information like transit schedules and travel times.

\subsub{D2. Level of Personalization}
Embedded information can be also categorized according to its degree of privacy and sensitivity, ranging from \textit{public} to \textit{personal} information.
For example, a user might augment their conversation through publicly available images like a picture of Rocky Mountain or personal photos from the user's own collection.
Other kinds of personal data, including information about past conversations or private search histories could be more useful, but entail trade-offs related to privacy, visibility, and trust --- both in the system and conversation partners. As M\"uller et al. highlight in their CloudBits work, including more private data may also have implications for the design and visibility of shared information~\cite{muller2017cloudbits}.

\subsub{D3. Modes of Communication}
Augmented conversation can support various modes of communication, including \textit{conversation, discussion, listening, presentation, and self-reflection}.
The modes of communication can be largely categorized as \textit{active} or \textit{passive}. In the active communication, the user actively engages with the conversation or reflection, such as in-person conversation, brainstorming discussion with a team, or self-reflection. 
On the other hand, augmented communication tools could also enhance the experience of passively listening to a conversation or discussion, including in larger meetings or lecture settings. 

\subsub{D4. Technology for Augmentation}
There are various ways to display conversation augmentations in the real world, including both \textit{on-body} devices and displays in the environment \textit{environment}.
These include a variety of immersive technologies, including \textit{head-mounted displays (HMDs), mobile AR on smartphones or tablets, and projection mapping, as well as more traditional displays and tabletops}.
Portable, on-body approaches like HMDs and mobile AR could allow users to augment conversations in many different environments, enabling more ubiquitous augmentation.
Personal devices like HMDs and mobile AR also allow more \textit{private and personal} context-driven referencing.
On the other hand, public displays and approaches like projection mapping could support \textit{public} augmented conversation, revealing consistent information to groups in settings like group brainstorms or in-person lectures.

\subsub{D5. Conversation Partners}
Augmented conversation approaches could apply to a variety of different conversation scenarios, including \textit{one-on-one} and \textit{group} discussions, as well as conversations with \textit{oneself} or with \textit{non-human agents}.
While one-on-one and group conversations have received the most discussion in the literature, visual augmentations for individual monologues (in which a person speaks aloud with no other audience) could create new opportunities for personal notetaking and ideation.
Augmented conversation with non-human agents, such as smartphones, smart speakers, IoT devices, and robots could also make it easier for people to interact with them without relying on mobile devices. For example, augmented conversation approaches could allow users to visually see keywords or concepts that emerge in podcasts or in exchanges with digital assistants.

\subsub{D6. Interaction Modalities}
There are a large number of possible ways to interact with augmentation conversation interfaces, including \textit{touch, gesture, controllers, voice commands, gaze or head movement, and proximity}.
The available interaction modalities are also related to which technology is used. For example, HMDs allow gaze and gestural interaction, while mobile AR does not. Similarly, projection mapping would be more suitable for touch or tangible interaction. 

\subsub{D7. Level of Interactivity}
Level of interactivity refers to how much active intervention an augmented conversation system requires. 
Systems with \textit{low interactivity} might show information related to the conversation automatically without requiring (or allowing) the viewer to direct what information is displayed or query further. On the other hand, \textit{high interactivity} systems could permit more active modification, organization, and querying.

\subsub{Summary of the Design Space}
Figure~\ref{fig:designspace} summarizes all of the dimensions that we discussed in this section, with the set of design options we explored in our \system prototype highlighted in blue.
While our current implementation considers only a subset of the possible designs, the broader design space highlights a numerous opportunities for future work. In fact, given recent advances in natural language processing and context-aware information retrieval, we expect that implementations of most of the possibilities we discuss are already possible.


\section{The \system System}
\begin{figure*}[ht]
\centering
\includegraphics[width=\linewidth]{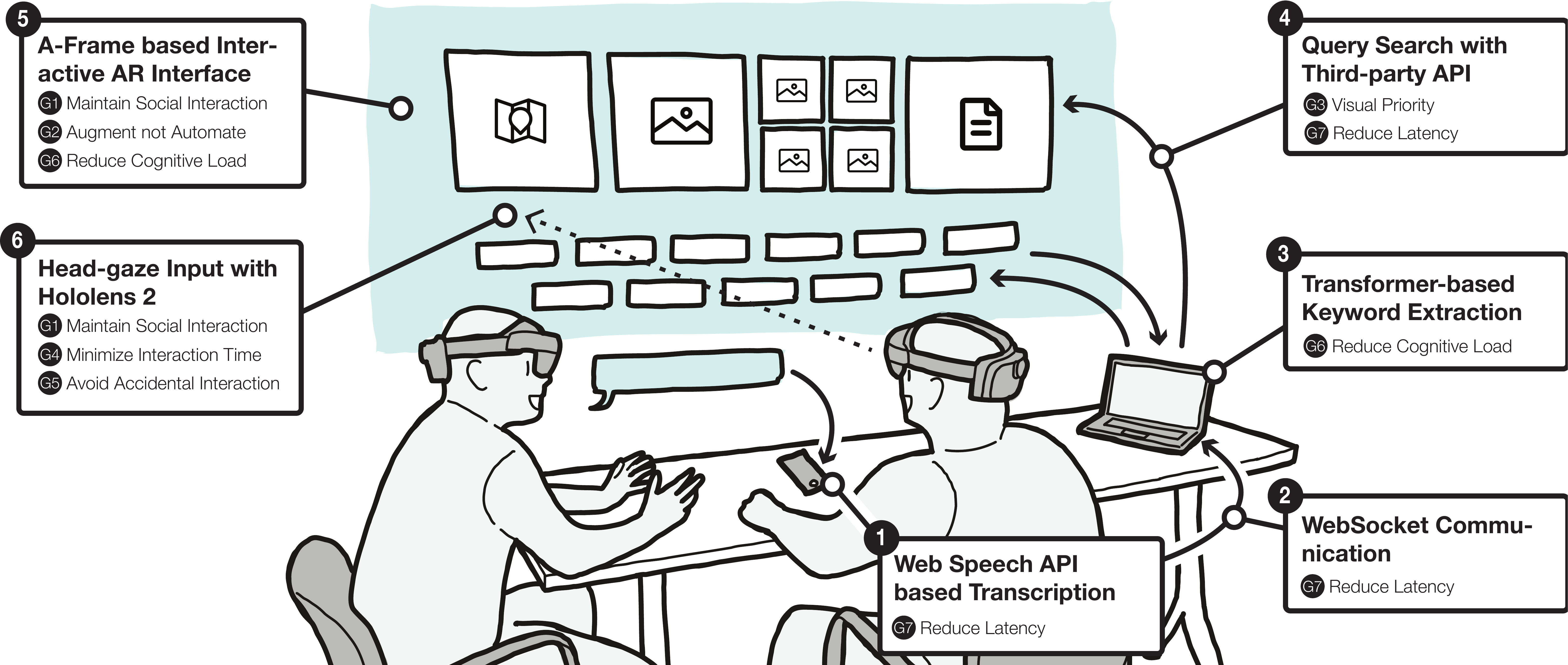}
\caption{Overview of the key features of our \system, each with corresponding design goals. }
\label{fig:authoring}
\end{figure*}

To demonstrate the concept of augmented conversation, we developed \system, an augmented reality system for on-the-fly context-driven referencing that instantiates many of the augmented conversation concepts described above. 
\system supports \textbf{real-time speech recognition}, \textbf{on-the-fly keyword extraction and referencing}, \textbf{contextual information retrieval}, \textbf{embedded visual references}, and \textbf{gaze-based interactions}. 
The prototype uses WebXR (A-Frame and React.js) for rendering, Web Speech API for real-time speech recognition, a Transformer-based neural network (for named-entity and keyword extraction), and various web-based APIs (Google Maps and Images, Wikipedia search, etc.) for context-driven keyword reference searches.
The prototype is then deployed on the HoloLens 2.
Figure~\ref{fig:teaser} illustrates the user interface and interactions with \system in an in-person conversation, while Figure~\ref{fig:authoring} shows the overall software structure. 

\subsection{Design Goals}
We developed our prototype using an iterative approach, starting from paper prototypes and low-fidelity experiments with a webcam and web browser. Gradually, we developed a functional tablet-based AR prototype before finally transitioning to a full head-mounted AR device, the HoloLens 2. 
At each stage, the authors tested each prototype themselves and informally recruited participants to test successive iterations in both scripted and unscripted conversations. During these design iterations, we formulated a series of design goals that guided the feature set and affordances of our final system.

\subsub{G1. Maintain Natural Social Interactions} 
The focus of social interactions (including conversations) is always the people, not the supplemental information, thus we should maintain natural conversational gestures such as eye contact. 
Thus, it is important to avoid any unnecessary distractions that might hinder the flow of the conversation. 
It became clear during the prototyping phase that AR itself created opportunities for distraction and the division of attention. For example, if the embedded information were overlaid on top of the social participant's face, it would erode the possibility of making eye contact. On the other hand, if the displayed information is very far away, it would force users to tilt their heads at awkward angles and unnecessarily prolong each interaction.
Therefore, it was imperative that we design the user interface such that the possibility of our system itself distracting the user be minimized by identifying bad interface designs and complex interaction techniques.

\subsub{G2. Augment Not Automate} 
In the prototyping phase, we have tested various different designs and approaches, including auto-generated word clouds and automated references.
We quickly noticed that the unnecessary automation increases the noise of the conversation, forcing the user to pay attention to unimportant information. Instead, a more flexible \textit{interactive system} that allows users to select which keywords to highlight and display contextual information about, would result in a less error-prone system for augmented conversations.

\subsub{G3. Encourage Visual Priority}
Through informal testing, we observed that textual representation of the keyword contextual references requires much more time to analyze and understand than visual representations. By the time users have read, analyze and understand the textual information, the conversation topic has changed or a break in the conversation was required to digest the information or users had to give up reading the textual representation at all to keep up with the conversation. Therefore, visual representations should be prioritized when possible.

\subsub{G4. Minimize Interaction Cost and Time}
When designing an augmented conversation interface, there are various potential interaction modalities. In AR, gesture, controller, or voice-based commands are the most common interface techniques. While expressive, these interactions often require more time and steps to interact with words and information than would be ideal. For example, gestural interactions, even a minimal air-tap, create an awkward behavior and atmosphere during a conversation. Voice commands are by nature impractical for our system since they'd require differentiating between keywords and voice commands, not to mention how bizarre and awkward it is in the middle of a conversation. Additionally, we observed that conversations are often fast-paced and always in real-time. This meant that the topic of conversation was fluid, often changing (sometimes quite randomly). Thus time is of the essence, and time spent interacting with or analyzing keywords is time not spent on social cues, focusing on the conversation or social participants. Thus, it is critical to minimize the time and effort required for users to interact with our system for keyword selection and lookup. This made gaze or head-movement-based the most attractive option due to its minimal interaction time and cost to maintain the flow of the conversation. 

\subsub{G5. Avoid Accidental Interaction}
Unexpected behaviors and interactions frustrate users and create the potential for both distraction and friction. Unfortunately, gaze- or head-movement-based interaction often causes accidental interactions or unexpected UI triggers, which undermines the user experience and delays critical reference information.
Thus a balance must be struck between minimal interaction and minimizing accidental interactions.
We alleviate this problem by confirming interactions by waiting for a couple of seconds or by adding visual feedback on the current state. That is, we use gaze-and-dwell as our interaction method.

\subsub{G6. Reduce the Cognitive Load for Content Consumption}
While prototyping, we modified the layout of a conversational transcript multiple times. Initially, we displayed the entire transcript, continuously updated as the conversation went on and highlighted identified keywords within the transcript. Not only was this not efficient, but it was time consuming and overwhelming to users. They didn't need an entire transcript nor was it intuitive to select keywords based on where in the transcript they occurred. Next, we modified the transcript to only display the last sentence. At this point, we already had to display the keywords elsewhere and informal user testing revealed that having a transcript wasn't really useful. Thus, our final prototype features no transcripts (as our visual priority design would recommend) but instead a list of keywords for users to select from. To reduce the cognitive load further, we colored keywords by category, red for people, blue for locations, purple for organizations and green for dates. Each category also has a corresponding icon. See figure 4 for an example. 

\subsub{G7. Minimize Latency for Real-time Augmentation}
Finally, latency is the key to a real-time conversation. 
Through our testing, it turns out this is a much more crucial factor than we originally anticipated. 
In a fast-paced real-world conversation, the topic can quickly change in real-time. 
Therefore, the system needs to have a small latency for \textit{every single aspect}, including speech recognition, keyword extraction, reference search, and visual rendering.
As we learned, for augmented conversation, latency can directly affect usability, thus minimizing the latency is the key to system design. 

\subsection{Implementation}
Based on the design goals, we implemented our final prototype built by WebAR for Microsoft HoloLens. 
Each design feature and functionality follows the design goals identified through the iterative design process. 
Our system features embedded references with visual priority, head-based interaction to minimize interaction cost, minimal content placement to reduce cognitive load and maintain social interaction, and reduce latency for real-time conversation. 

\subsub{Corresponding Design Space}
As we discussed in Section 3, there are many possible implementation approaches, but in our system, we specifically focused on \textbf{HMD-based approach (D4)} using Microsoft HoloLens 2 for the ubiquity of the augmentation, given the prediction of HMD will be widely available in the next decades. 
For other design spaces, we covered \textbf{most of the types of associated information (D1)} but focused on \textbf{non-personalized information (D2)} due to the difficulty of implementing personalized information for user testing. 
For interaction, we focus on \textbf{gaze- or head-based interaction (D6)} and \textbf{low level of interactivity (D7)} informed by the design goals and user-centered design process.
While our system itself can be used for various modes of communication (D3) and use cases (D5), we focus on our use cases mostly centered around \textbf{in-person conversation (D3 and D5)}, which is our focus in the user evaluation study. 
In the following subsections, we describe the implementation details.

\subsubsection{Real-time Speech Recognition using the Web Speech API}
First, the system uses the Web Speech API for real-time speech recognition and transcription. 
Web Speech API is a widely available speech recognition engine that can be used in a browser.
The reason why we used Web Speech API is to \textbf{reduce the latency (G7)}. 
We originally tested with relatively higher-quality server-based Google or Microsoft speech recognition API, but we noticed that this API-based approach often courses higher latency. Thus, we use Web Speech API. 
Microsoft HoloLens 2 has HoloLens Edge browser, but Web Speech API is not yet available in HoloLens Edge browser, thus we place the Android phone next to the user and use Google Chrome-based Web Speech API instead.

Web Speech API can have several options such as \textit{continuous} mode, in which we can decide whether or not to use a speech sentence before the speaker finishes talking.
We enable continuous mode and mainly use intermediate results for lower latency.
In a regular conversation with 110 words/min, our Web Speech API implementation recognizes, transcribes, and send the text at 360 times/min for continuous mode and 10 times/min for single mode.

\subsubsection{Synchronous Communication through WebSocket}
In our system architecture, a web browser (Android's Google Chrome) transcribes the speech to text and sends the detected text to a Node.js web server on MacBook Pro through the WebSocket protocol.
WebSocket protocol allows faster synchronous communication to \textbf{reduce the latency (G7)} than the other protocol like TCP/IP communication.
Once the Node.js server detects the keywords through the above pipeline, it broadcasts the result to all connected browsers (Microsoft HoloLens Edge browser) in JSON format through the WebSocket protocol.
The server-side machine (MacBook Pro), runs a Node.js server and Python script. Whenever the server receives the newly transcribed text, it runs and gets the result through an efficient inter-process communication based on Node PythonShell. The overall latency from sending the transcribed text to receiving the result in a client-side browser is under 10ms. 

\begin{figure}[tb]
\centering
\includegraphics[width=1\linewidth]{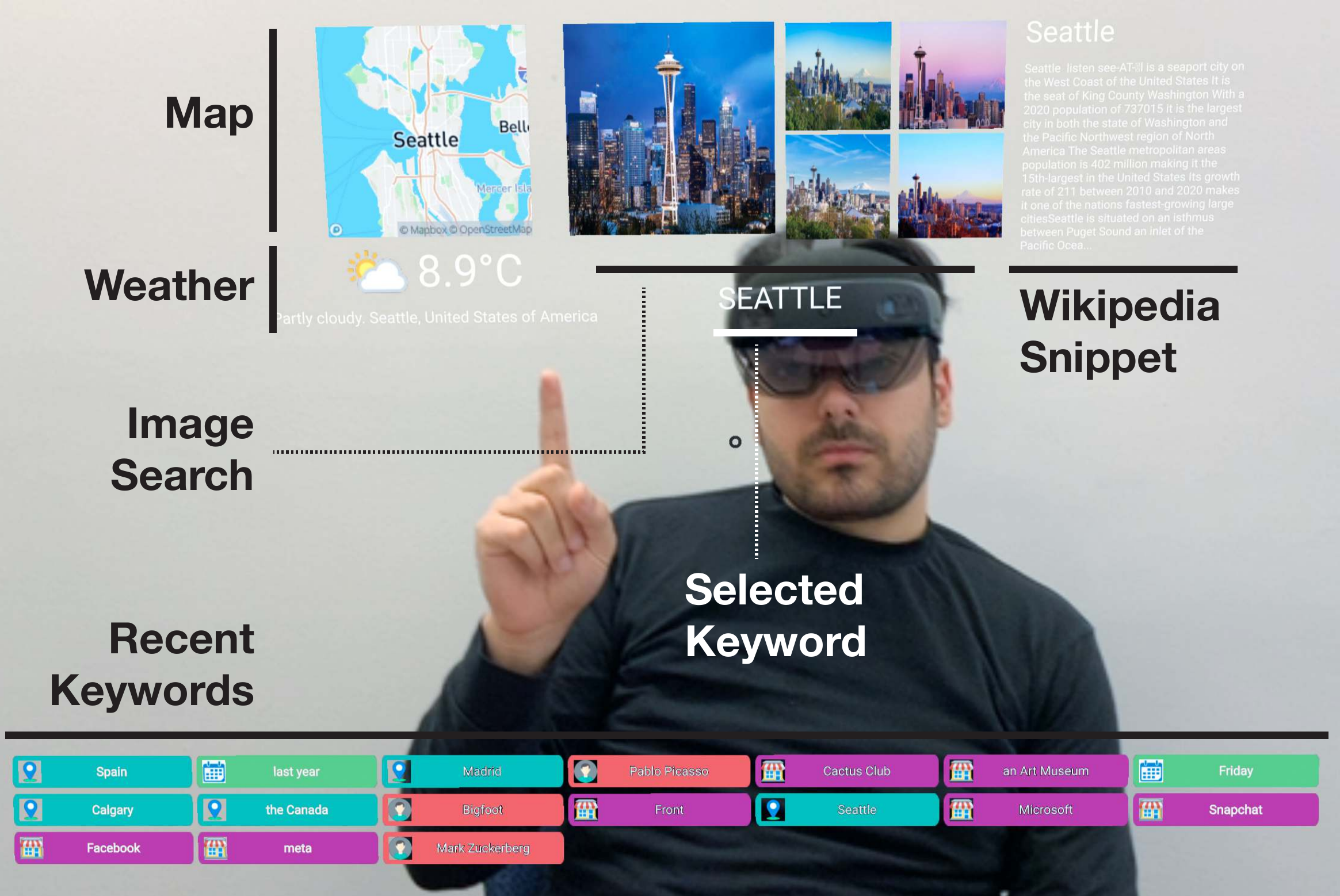}
\caption{The \system system as seen from the point of view of a user wearing a HoloLens 2.}
\label{}
\end{figure}

\begin{figure*}[tb]
\centering
\includegraphics[width=\linewidth]{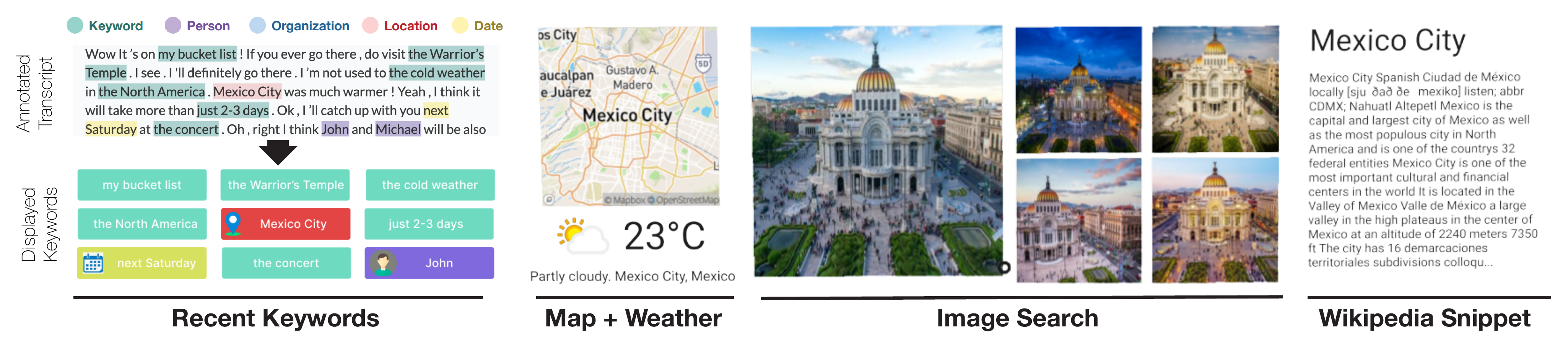}
\caption{Closeup views of the visual reference types implemented in \system including keywords, maps, weather, image search results, and Wikipedia snippets. Keywords are extracted from the live transcript (upper left) and labeled by category (organization, location, date, and person). The transcript itself is not shown in the AR view. }
\label{}
\end{figure*}

\subsubsection{Real-time Keyword Extraction through Transformer-Based Natural Language Processing}
Second, the system extracts keywords and categorizes the types of keywords based on natural language processing.
The reason why we used keyword extraction is to \textbf{reduce the cognitive load for content consumption (G6)}.
We originally let users directly select words from the transcript. However this causes several issues, since the transcript flow was often extremely fast, which caused the amount of content to quickly become overwhelming.
While we tested several keyword extraction engines, we ultimately used Spacy3, a Transformer-based deep neural network pipeline built with Python and Cython.
We use a pre-trained model based on Spacy's tokenizer and word embeddings model (en\_core\_web\_sm).

In our NLP pipeline, we first detect noun phrases and combine words (for example, we treat \textit{``Human Computer Interaction''} as one word, instead of \textit{``Human''}, \textit{``Computer''}, and \textit{``Interaction''}). 
Then, we detect the named entity for each word to classify the word to different categories. 
For named-entity categories, we have \textit{organization}, \textit{location}, \textit{person}, \textit{date}, and \textit{numerical values} (for example detect \textit{``Google''} as \textit{``organization''} and \textit{``New York''} as \textit{``location''}).
Finally, we extract and rank keywords using a simple TextRank algorithm~\cite{mihalcea2004textrank}, which extracts key phrases based on a graph algorithm that is independent of a specific natural language and does not require domain knowledge. 
We use the phrase which has a positive TextRank value as extracted key phrases.
The Transformer-based keyword extraction pipeline runs with Python 3.7 on MacBook Pro 16inch Intel Core i7 CPU, 16GB RAM.
Our keyword extraction pipeline works in real-time with an average of 684 words per second on CPU.
The response time will quickly increase based on the number of words, thus we run this extraction pipeline separately for each intermediate result, rather than all of the accumulated text.

\subsubsection{Query Search through Third-party API}
When the user selects a word, the system also provides a reference related to the information based on \textbf{visual priority (G2)}.
For associated information, we implemented \textbf{\textit{image search}} based on the Google and DuckDuckGo Image APIs, \textbf{\textit{maps}} based on MapBox API, \textbf{\textit{search results and news articles}} based on Google and DuckDuckGo Custom Search APIs, \textbf{\textit{weather}} based on the Open Weather Map API, 
and \textbf{\textit{text descriptions}} based on Wikipedia API. 
All of the API requests are completed either through client-side JavaScript or server-side Node.js code.
For all words, we provide visual references based on \textit{images} and \textit{search results}. 
To provide richer context-driven references for entities, we use a custom query search based on the named entity category identified through the NLP pipeline.
For example, for location-related keywords we provide \textit{maps} and \textit{weather}, for proper nouns we provide \textit{Wikipedia} information, for date-related information we show a \textit{calendar}, etc.
To \textbf{reduce the latency (G7)}, we pre-process all of the extracted keywords in the background to obtain the associated information (only images and search results) before the user interaction happens, so that the system can show this associated information immediately when the user selects it.

\subsubsection{Embedded Visual Reference Rendering using A-Frame, Three.js, and HTML Canvas}
All of the processed results including extracted keywords and associated information are shown through embedded graphics in AR.
For rendering, we implemented with A-Frame and Three.js, embedded in the 3D space. 
Since most of the information can be shown as 2D objects, we use Three.js Canvas Textures to show them as an interactive HTML element built with React.js and Semantic UI.

Our user interface design is followed by the design goals.
First, we implemented a JavaScript-based prototype that renders the interface, along with all the associated information panels close to the speaker's face. This allows the user to maintain the eye-contact while showing the information at a minimal distance from the other user.
This helps the user to 
\textbf{maintain natural social interaction (G1)}.
Second, we build our system as an interactive interface so that the user can select the required information by themselves through interaction, based on \textbf{augmentation rather than automation (G2)} principle.
Our interface basically consists of a transcribed text, keyword list, and associated information. The user mainly interacts with the keyword list, compared to the raw transcribed text to reduce the noise and easily let the user identifies the topic.
This can \textbf{reduce the cognitive load (G6)} for the user.
Also, we only have one main associated information panel so that the user can only focus on one thing, while the user can quickly access the other information through the accompanied history view on the side.


\subsubsection{Head-Gaze Interaction through Hololens 2}
We use Microsoft Hololens 2 for AR rendering devices.
To \textbf{minimize interaction time (G4)}, we use head-gaze as a primary interaction modality.
While Hololens 2 provides both eye-gaze and head-gaze input, in the WebXR framework, currently head-gaze input is only available.
Therefore, we place a cursor in the center of the screen and let the user selects the word by moving her head.
To \textbf{maintain natural social interaction (G1)}, we do not require a common selection technique like air tap.
Instead, whenever the cursor hovers the clickable object, the system recognizes it as a selection input.
However, this native implementation often causes unexpected input. 
To \textbf{avoid accidental interaction (G5)}, we also implemented a time-based trigger and visual indicator, so that the system only confirms the selection when the user pauses for at least 250 ms.


\section{Evaluation}
We validated our system via an exploratory usability study. The goal of this study was to evaluate the usability of our prototype and to identify limitations or opportunities for future improvements. 

\subsection{Method and Participants}
We recruited 13 participants (9 female / 4 male, ages 18-27) to each take part in a 50 minute in-person session in which they used our prototype during three topic-driven conversations with a member of the study team. 
We first introduced the concept of augmented conversation and gave 5 minute live demo of the \system system. 
Then, we asked the participants to take part in three 5-minute conversations with one of the experimenters. Each conversation focused on a specific topic (vacation, career, or family). To ensure that all participants had a similar experience, the experimenter used a consistent script to guide their responses, but allowed the participant to direct the conversation and introduce new topics. All participants were fluent English speakers (9 were native English speakers while the rest spoke English as their second language) and conducted all conversations in English. 

During the conversation, both the participant and the experimenter wore a Microsoft HoloLens 2 and participants had access to the \system prototype for the entirety of their conversation. 
After the session, we asked the participants to give feedback about the interface and their experience through an online questionnaire. Finally, we conducted a short semi-structured interview to capture qualitative feedback and asked participants to complete a short recall quiz designed to assess their attention to the conversation. The quiz contained 10 questions with answers drawn from the experimenter's conversation script. 
We recorded all participant interactions both through the participant's HoloLens and via an external video camera.
Each participant was compensated \$10 CAD.

\subsection{Results} 
Overall, the participants responded very positively to their experience with the prototype. 
When we asked 7-point Likert-scale questions about the experience (1=\textit{Strongly Disagree}$\leftrightarrow$ 7=\textit{Strongly Agree}), participants reported that they found the overall system helpful and intuitive ($\mu$=5, $\sigma$=1.15). Participants also agreed that the referencing features were very helpful and directly relevant to the conversation ($\mu$=6, $\sigma$=0.91). 
Moreover, all participants agreed that the approach had great potential and expressed interest in using augmented conversation tools in the future. Broadly, our observations from the study highlighted the relevance of visual references, as well as trade-offs related to distraction and system efficiency.

\subsubsection{Content Relevance}
Participants found the references very useful, with almost all reporting that the visual references were directly relevant to the conversation ($\mu$=6, $\sigma$=1.15). 
For example, P11 noted that \textit{``All the references that did pop up were directly related to the main ideas of what was said.''}, while P8 added that \textit{``The keywords were really helpful because looking back at them helped me remember the key topics of conversation so far.''} (P8). 

Images were the most popular visual reference ($\mu$=5.5, $\sigma$=0.77), followed by maps ($\mu$=5.1, $\sigma$=1.14) and text ($\mu$=4.8, $\sigma$=1.46). The majority of the people neither liked, nor disliked the weather references ($\mu$=4.38, $\sigma$=1.04). However, 10 out of 13 participants preferred highly visual features over text --- with P11 noting specifically that they ``could not read [text] and focus/ participate in the conversation at the same time'' but could take in pictures ``at a glance''.  


\subsubsection{Visual Augmentation and Distraction}
Participants also generally agreed that the AR system helped reduce distraction when compared to smartphone search, reporting that the dynamic rendering of references made it easier to stay engaged in the conversation --- providing visual anchors and history that helped them stay connected to the conversation rather than diverging from it. 
As one participant (P8) remarked, \textit{``I found that the AR references helped me keep track of the main topics during the discussion. In fact, I found the AR references less distracting than using my phone when trying to maintain a meaningful conversation that goes both ways.''} Even for participants who became distracted reported that  visual priority (G2) of the references helped them to recall information even after the conversation. P11 emphasized this point, stating that \textit{``since I am a visual learner, I was able to recall the main points because I had \textbf{seen} the keyword [emphasis added]''}. P8 reported a similar experience, noting that ``\textit{``the AR references were useful subheadings that helped me keep track of the main topics in discussion''}.
%
%
%
However, focusing on conversations and references at the same time still sometimes caused divided attention. P11 in particular noted that the \textit{``displayed images and keywords were slightly distracting''}, but also emphasized that the visual references' placement in their peripheral vision helped them from being too visually overwhelming. 

\subsubsection{Interaction and System Efficiency}
In terms of the interaction, the participants had a mixed impression of the system, highlighting how the success of augmented conversation systems may depend heavily on their ability to surface information quickly.
However, despite our focus on minimizing latency, the complexity of the referencing pipeline often meant that keyword gaze interactions took a long time to trigger visual references. For example, P12 noted that \textit{``most of the keywords popped up pretty quickly, but some keywords popped up at a later time''}. 
Similarly, P3 mentioned that \textit{``keywords took too long to open. I felt like we switched topics by the time it opened.''}. 

However, participants liked the gaze-based interaction, with P3 noting that \textit{``It's like searching things up with your eyes''}. 
P6 also highlighted how gaze-based interactions helped focus their attention stating \textit{``it felt just like a normal conversation, as though your point-of-view became an extension of your phone [in terms of on-the-fly referencing] while avoiding potential distractions like viewing 40 irrelevant notifications''}.
\section{Discussion and Future Work}

While our initial exploration focused primarily on one-on-one conversations, augmented conversation tools hold promise for a variety of other settings, including education (taking lecture notes, attending conferences, group studying, etc.) and professional (attending meetings, presenting, brainstorming sessions, etc.) scenarios. 
As our transcription pipeline can help dynamically create meeting logs or lecture notes, it could be a highly useful tool not only for real-time referencing but also for documenting events for future reference. As P9 mentioned, \textit{``I can see that if I was being taught about the synthesis of aspirin that I could look up the different steps involved and take notes based on the flow of the conversation``}. 
For educational settings, it could be also interesting to explore different technical approaches and richer interactions. Participants noted that future augmented conversations could help in creating an outline in lectures and conferences (P8), break down subject matter in steps for note taking (P9), help naturally search information in conversation (P2), help them easily take notes while listening to lectures (P4), and provide visual referencing to spoken words in lectures (P11).
Moreover, in the classroom, it might make more sense to use a projection mapping approach, similar to HoloBoard~\cite{gong2021holoboard}, or projecting mobile AR screens like RealitySketch~\cite{suzuki2020realitysketch} rather than asking all of the students to wear HMDs.
Going forward, we are especially interested in designing, deploying, and evaluating the system in classroom or meeting settings. 
%
While our work mostly focuses on augmenting everyday conversations, we also see so potential for accessibility assistance both for deaf and hard-of-hearing people as well as people with visual impairments. Applications in this domain might building on the kinds of captioning tools developed by accessibility researchers~\cite{mosbah2006speech, peng2018speechbubbles, olwal2020wearable, tu2020conversational, miller2017use, schipper2017caption, findlater2019deaf, jain2018exploring, jain2015head, jain2018towards} while adding a range of additional audio-visual referencing. 
\section{Conclusion}
We present RealityChat, a system that supports co-located in-person conversations via embedded speech-driven on-the-fly referencing in augmented reality. With the help of AR, RealityChat provides relevant visual references in real-time, based on keywords extracted automatically from the spoken conversation. Our study confirms that our system helps in reducing distraction in conversations while providing highly useful and relevant information via simple interactions.  Moreover, our experiences augmenting day-to-day conversations highlight broader opportunities for augmented conversation systems. Future tools have the potential to integrate context-based personal information, provide accessibility assistance, and provide value in a variety of educational and professional use cases.


\balance
\bibliographystyle{ACM-Reference-Format}
\bibliography{references}

\end{document}